\documentclass[twocolumn,showpacs,preprintnumbers,amsmath,amssymb]{revtex4}

\usepackage{epsfig}
\usepackage{graphicx}
\usepackage{dcolumn}
\usepackage{bm}

\begin{document}


\title{Huge metastability in high T$_c$ superconductors induced by parallel magnetic field}
\author{R. G. Dias}
\affiliation{Departamento de F\'{\i}sica, Universidade de Aveiro, \\
3810 Aveiro, Portugal.}

\author{J. A. Silva}
\affiliation{Instituto Polit\'ecnico de Portalegre, \\
7301 Portalegre Codex, Portugal.}
\date{\today}


\begin{abstract}
We present a study of the temperature-magnetic field phase diagram of
homogeneous and inhomogeneous
superconductivity in the case of a quasi-two-dimensional superconductor
with an extended saddle point in the energy dispersion under a parallel magnetic field.
At low temperature, a huge metastability region appears, limited above by a
steep superheating critical field (H$_{sh}$) and below by a strongly reentrant
supercooling field (H$_{sc}$).
We show that the Pauli limit (H$_{p}$)  for the upper critical magnetic field
is strongly enhanced due to the presence of the Van Hove singularity in the density of states.
The formation of a non-uniform superconducting state is predicted to be very unlikely.
\end{abstract}


\maketitle
The theoretical phase diagram of two-dimensional isotropic superconductors
under parallel magnetic fields
shows  distinctive features such as a tricritical point where a high temperature second order phase transition
gives way to a low temperature first order transition \cite{maki,shim} and consequently to
a region of metastability. Furthermore, in such systems,
superconductivity may persist at magnetic fields higher
than the first order critical field if the formation of an
inhomogeneous superconducting phase, i.e., the so-called Fulde-Ferrel (FF) phase,\cite{shim,fuld,lark}
occurs.

Superconducting copper-oxides have strong quasi-two-dimensional character, but the energy dispersion of the
CuO$_2$ planes is clearly anisotropic and in particular, it contains extended saddle points
as documented by angle resolved photoemission experiments.\cite{gof}
Extended saddle points have also been found in the case of the two-dimensional Hubbard model
by Assaad and Imada \cite{imad}  using the Quantum Monte Carlo technique.
Such saddle points lead to strong Van Hove singularities (VHS) in the density of states
and recently, it has been suggested \cite{dias} that such singularities could provide an explanation for the
strange out-of-plane upper critical field  of the High-T$_c$ superconductors.\cite{mck}
In this paper, we study  the phase diagram of a two-dimensional superconductor
with an extended saddle point in the energy dispersion under parallel magnetic field.


In a strictly two-dimensional superconductor under an in-plane magnetic field,
orbital frustration due to magnetic field
is not present and the upper critical field is determined by Zeeman pair breaking.
Therefore, we consider the Hamiltonian
\begin{eqnarray}
        H &=& \sum_{{\bm k} \sigma} \xi_{{\bm k} \sigma} a^\dagger_{{\bm k} \sigma} a_{{\bm k} \sigma}   \\
        && -{V \over S} \sum_{{\bm k}_1 {\bm k}_2 {\bm k}_3 {\bm k}_4}
        \delta_{{\bm k}_1-{\bm k}_2,{\bm k}_4-{\bm k}_3}
        a^\dagger_{{\bm k}_1+{\bm q}/2 \uparrow} a^\dagger_{-{\bm k}_2+{\bm q}/2 \downarrow} \nonumber \\
        && \times a_{-{\bm k}_3+{\bm q}/2 \downarrow} a_{{\bm k}_4+{\bm q}/2 \uparrow} \nonumber
\end{eqnarray}
with $ \xi_{{\bm k} \sigma}=\xi_{\bm k} -\sigma h$, $\xi_{\bm k}=\epsilon_{\bm k}-\mu$ and $h=\mu_B H$
and where $V$, $\mu$, $H$, $S$ and $\mu_B$ are respectively the attractive interaction constant,
the chemical potential, the magnetic field, the area of the system and the Bohr magneton. The ${\bm k}$
sums in the interaction term follow the usual BCS restrictions.

S-wave symmetry is assumed throughout the paper, for simplicity. The effects of d-wave symmetry in the case
of a superconductor with constant density of states have been addressed by several authors. \cite{yang}
It leads to visible modifications in the phase diagram for inhomogeneous superconductivity with, in particular,
the appearance of a low temperature kink in the phase boundary between the FF phase and the normal phase.
However, the phase diagram for homogeneous superconductivity remains very similar to that of an s-wave superconductor.

Minimizing the free energy \cite{rick,shim} or, equivalently,
following a BCS mean field approach in the case of an homogeneous gap function,
$\Delta=-VS^{-1}  \sum_{{\bm k}´} \langle
a_{-{\bm k}´ \downarrow} a_{{\bm k}´ \uparrow} \rangle $,
one obtains the two-dimensional gap equation
$
        1 = VS^{-1} \sum_{\bm p} [1 -f(\xi^{\Delta}_{\bm p \uparrow})
        -f(\xi^{\Delta}_{\bm p \downarrow})] ( 2 \xi^{\Delta}_{\bm p\uparrow})^{-1}
$
with
$ \xi^{\Delta}_{{\bm k} \sigma} = \xi^{\Delta}_{\bm k} -\sigma h$,
$\xi^{\Delta}_{\bm k} = \sqrt{\xi^2_{{\bm k}}+\Delta^2}$
and where  $f(x)$ is the Fermi distribution function.
Taking the limit $\Delta \rightarrow 0$,
one obtains
\begin{equation}
         {1 \over V} = \int_0^{\omega_D} d\xi {N(\xi) \over 2 \xi}
         \left[ \tanh {\xi - h  \over 2t} +
         \tanh {\xi + h  \over 2t} \right]
         \label{eq:gapequation1}
\end{equation}
where $\omega_D$ is the usual frequency cutoff and $t=k_B T$.
From this equation, one extracts the temperature dependence of the critical field (H$_{sc}$)
that induces the second order phase transition
from the homogeneous superconducting state to the normal state.
Below a certain temperature (the tricritical point temperature), the phase transition
becomes of first order and the field
given by the previous equation becomes a supercooling field above which the normal state is a local
minimum of the free energy.
Note that this critical field has no dependence on the Fermi surface shape, only density of states dependence.

In a two-dimensional system, a Van Hove singularity in the
density of states results usually from the presence of a saddle
point in the energy dispersion $\epsilon({\bm k})$.
In the case of a simple
saddle point, $\epsilon({\bm q}) \sim q_x^2 - q_y^2$, one has a logarithmic
singularity in the density of states. In the case of an extended
saddle point, $\epsilon({\bm q}) \sim \vert q_x \vert^n - \vert q_y \vert^m$,
with higher powers and
this leads to a power-law divergence in  the density
of states $N(\epsilon) \sim (\epsilon-\epsilon_{vh} )^{-\alpha}$ with $\alpha=1-{1\over n}-{1\over m}$.
Besides two-dimensional systems, power-law divergences are present in one dimensional systems.
In a one-dimensional system, a $q^2$ dispersion leads to
a inverse square root divergence in the density of states.
In the following, we will assume that the VHS is pinned at the Fermi level as observed in photoemission
experiments in the copper-oxides \cite{gof}
and also indicated by theoretical studies of correlated two-dimensional models.\cite{gon}

Assuming $N(\epsilon) =N_o \vert \epsilon-\epsilon_{vh} \vert^{-\alpha}$,
in the weak coupling limit, Eq.~\ref{eq:gapequation1} can be rewritten as
\begin{equation}
         {\alpha \over g}= {1 \over (2t)^{\alpha}} G \left( {h \over 2t} \right)
         \label{eq:resultzeeman}
\end{equation}
with $g=VN_o/2$ and
\begin{equation}
          G(x)=\int_0^\infty d\xi {\xi^{-\alpha} \over  \alpha} \left[ {1 \over  \cosh^2 (\xi+x)} +
          {1 \over  \cosh^2 (\xi-x)} \right].
\end{equation}
The upper limit of the integral in Eq.~\ref{eq:gapequation1} was extended to infinity due to the fast decay
of the integrand and therefore, in the weak coupling limit,
the supercooling field has no dependence on the frequency cutoff.
The zero field critical temperature obtained from Eq.~\ref{eq:resultzeeman} is given by
$
          2t_{c0} \sim [  g/(\alpha- \alpha^2)
          ]^{1/\alpha}
$
which implies the well known enhancement of critical temperature which motivated the Van Hove scenario
of high $T_c$ cuprate superconductors.
The zero temperature supercooling  field ($h_{sc,0}$) is given by
$
          h_{sc,0} = (2 g/\alpha)^{1/\alpha}
$.\cite{remark}
Therefore, $ h_{sc,0} \sim (1- \alpha)^{1/\alpha} t_{c0}$ and one has a much
larger enhancement of $t_{c0}$ than that of $h_{sc,0}$ in the limit of
very strong VHS, $\alpha \rightarrow 1$.
The reason behind the different enhancements is reflected in the fact that for fixed finite temperature
and zero magnetic field, the pairing susceptibility diverges in the limit $\alpha \rightarrow 1$,
but it  does not diverge for fixed finite magnetic field
and zero temperature (note that $\tanh(x)$ is linear for small $x$).

\begin{figure}[tbp]
       \begin{center}
       \leavevmode
       \hbox{%
       \epsfxsize 3.0in   \epsfbox{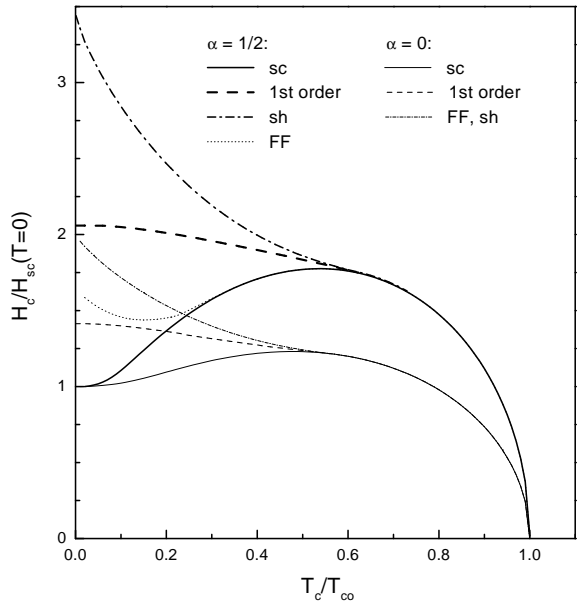}}
       \end{center}
       \caption{The phase diagram of a paramagnetically limited two-dimensional superconductor
       with constant density of states ($\alpha=0$) and with a power-law divergence in the density of states
       ($\alpha=1/2$) pinned at the Fermi level.
       The reentrant behavior of the supercooling field  $H_{sc}$ observed in BCS superconductors
       is strongly enhanced if a VHS is present in the DOS.}
       \label{fig:qual}
\end{figure}

In Fig.~\ref{fig:qual}, the temperature  dependence of the reduced upper critical field (or supercooling field)
obtained numerically from Eq.~\ref{eq:gapequation1} is displayed for both a superconductor with a VHS
and a superconductor with constant density of states (BCS superconductor).
The VHS exponent $\alpha=1/2$ is obtained in the case of a quadratic one-dimensional energy
dispersion or for example, in the case of an extended saddle point with quartic dispersion as found in
the two-dimensional Hubbard model.\cite{imad}
One observes in Fig.~\ref{fig:qual} that the maximum second order critical field is not reached
at zero temperature, but at an intermediate temperature. This
reentrant behavior for the supercooling field
is known in BCS superconductors,\cite{maki} and it has been recently observed in
thin aluminum films.\cite{prl-al}
In the case of a  VHS superconductor, this maximum is enhanced relatively to the
zero temperature supercooling field and the reentrance becomes more pronounced.

The strong reentrant behavior of h$_{sc}$ can be explained in the following way.
At zero temperature, the Fermi surface splitting in the normal ground state
due to the Zeeman coupling
creates a no-pairing or blocking region  around the zero field Fermi surface since, in order to contribute
to the formation of a Cooper pair of momentum ${\bm q}$, two states with opposite spins
and of momenta ${\bm k}+{\bm q}$ and $-{\bm k}$ must be either both empty or both occupied
at zero temperature.
In the case of homogeneous superconductivity, Cooper pairs have zero total momentum, $q=0$.
At finite temperature, thermal excitations provide
low energy pairing possibilities within this no-pairing region and therefore,
the pairing susceptibility grows with temperature.
In the presence  of a VHS at the Fermi level, the number of these thermal excitations
is much larger due to the strong splitting of the Fermi surface
in the saddle point regions and therefore, this effect is more pronounced.


Given a certain value of the superconducting gap, the free energy difference between
the superconducting and the normal state can be determined from \cite{maki}
\begin{equation}
    F_s(T,H)-F_n(T,H)=\int_0^\Delta {d(1/V) \over d \Delta'} \Delta'^2 d \Delta'
    \label{eq:freeenergy}
\end{equation}
The transition to the normal state with variation of field or temperature occurs
when the zero gap local extreme of the free energy becomes the absolute minimum.
If the finite gap local minimum of the superconducting phase converges to the normal state
local extreme as the transition is approached, the transition is of second order,
otherwise it is of first order
and there is a region of metastability limited below by the supercooling field $h_{sc}$
and above by  the superheating field $h_{sh}$. The superheating field is the highest field
for which there is still a finite
gap solution of Eq.~\ref{eq:gapequation1}.\cite{maki} It can be shown that
below this field, there is a finite gap local minimum in the free energy.
The  first order critical field $h_{1}$ and the  superheating field determined numerically from
Eqs.~\ref{eq:gapequation1} and \ref{eq:freeenergy} are shown in Fig.~\ref{fig:qual}.
At zero temperature, the superheating field is just the zero temperature gap function,
$h_{sh,0}=\Delta_0 = [A(\alpha)g]^{1/\alpha} \sim t_{c0}$, with $A(\alpha)=2\pi \text{cosec} [(1-\alpha)\pi] P_{-\alpha}(0)$
where $P_\nu(x)$ is the Legendre function of the first kind.
Consequently,  $ h_{sc,0}  \sim (1- \alpha)^{1/\alpha} h_{sh,0}$
and  there is a huge metastability region.
In Fig.~\ref{fig:qual}, it is also apparent that the reduced temperature of the tricritical point
is slightly larger in the case of the VHS superconductor.

At zero temperature, the free energy difference is given by
\begin{equation}
    \Delta F(0,H)={N_o \over 4-2 \alpha } \left[ {4 \over 1-\alpha}\, h^{2-\alpha}
    - \alpha A(\alpha) \, \Delta^{2-\alpha} \right].
\end{equation}
The zero temperature critical field associated with the first order transition is usually denominated
by Pauli limit or Chandrasekhar-Clogston limit \cite{Shan}. This field is easily extracted from the previous result,
$
    h_p =\{  \alpha(1-\alpha)
    A(\alpha) /4 \}^{1 \over 2-\alpha} \Delta_0
$
and therefore, this limit to superconductivity is as strongly enhanced
as the zero temperature superconducting gap
due to the presence of the VHS. For $\alpha =0$, one recovers the well known
result, $h_p=\Delta_0/\sqrt{2}$.\cite{maki} In the limit $\alpha \rightarrow 1$,
$h_p \rightarrow \Delta_0/2$.
Note that recent experiments indicate that the in-plane
upper critical field  in the copper-oxides exceeds considerably the BCS Pauli limit.\cite{exp1}


The previous results are independent of the shape of the Fermi surface. Now, we will consider
finite ${\bm q}$ solutions of the gap equation, that is, we will search for a Fulde-Ferrel phase
in the phase diagram.
Following the  BCS mean field approach in the case of an inhomogeneous gap function,
$\Delta_{\bm q}=-VS^{-1}  \sum_{{\bm k}´} \langle
a_{-{\bm k}´-{\bm q}´ \downarrow} a_{{\bm k}´ \uparrow} \rangle $,
and taking the limit $\Delta_{\bm q} \rightarrow 0$,
one obtains the following two-dimensional gap equation,\cite{shim}
\begin{equation}
         1 = {V \over S } \sum_{\bm p} {1 -f(\xi_{\bm p+ \bm q/2 \uparrow})
         -f(\xi_{\bm p - \bm q/2 \downarrow})\over \xi_{\bm p+ \bm q/2 \uparrow}
         +\xi_{\bm p - \bm q/2 \downarrow}}  .
         \label{eq:gapequation2}
\end{equation}
For a given temperature, the FF critical field is determined by searching
for the highest field solution of this equation
for any value of ${\bm q}$.

It is well known that the Fulde-Ferrel state becomes enhanced in the presence of
Fermi surface nesting.\cite{dupu} One might imagine that a VHS pinned at the Fermi level
could provide a similar nesting effect in the Fulde-Ferrel state.
However, one should be aware of a difference: while Fermi surface
nesting does not enhance homogenous superconductivity, in the case of VHS nesting it is considerably  enhanced.
It is therefore possible that  the Fulde-Ferrel region of the phase diagram is
narrower or that it even does not exist.
We will show that at zero temperature and finite magnetic field,
saddle points lead in fact to very poor Fermi surface nesting and
indeed the Fulde-Ferrel region of the phase diagram is
absent in the case of an extended saddle point.
Furthermore, this phase remains absent even if one improves the nesting property by
considering an isotropic Fermi surface
with vanishing Fermi velocity.

Let us consider the latter first.
In order for an isotropic energy dispersion to have  an VHS in the density of states,
it must be of the form
$
     \xi_{\bm k}=\epsilon_{\bm k}-\epsilon_{vh}=a \cdot \text{sign}(q) \vert q\vert^b
$
where ${q}=k-k_{vh}$. Again, we assume $k_F=k_{vh}$.
The density of states for the above model is $N(\epsilon)\sim
a^{-1/b} b^{-1} \xi^{{1\over b}-1}$ and therefore
the inhomogeneous gap equation can be rewritten as
\begin{equation}
         1= V \int_{0}^{\omega_D}d\xi N(\xi) \int_0^\pi {d\theta \over \pi}
         {\tanh \left( {\xi^+ \over 2t} \right)+
         \tanh \left( {\xi^- \over 2t} \right) \over \xi^+ +
         \xi^- }
         \label{eq:gapequation3}
\end{equation}
with
$
         \xi^\pm=[\vert \xi \vert^{1/b} \text{sgn} (\xi) \pm 1/2 q \cos \theta]^b \pm h
$
where we have neglected a $q^2/k$ term since $q \ll k_F$ and where $[x]^b$ should be understood
as $\text{sgn} (x) \vert x\vert^b$. In order to keep the expressions simpler, we have considered $a=1$.
Note that in the one-dimensional case, the angle integration would be absent.

\begin{figure}[tbp]
       \begin{center}
       \leavevmode
       \hbox{%
       \epsfxsize 3.5in   \epsfbox{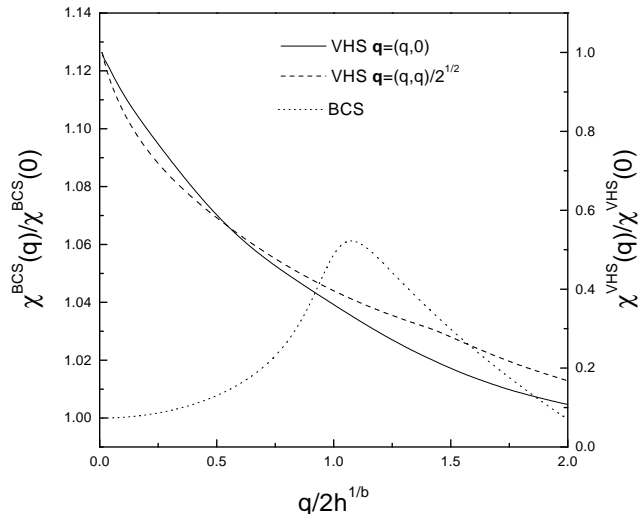}}
       \end{center}
       \caption{ The pairing susceptibility as a function of the renormalized
       pair momentum $q/(2h^{1/b})$
       for a BCS superconductor and a superconductor with an extended VHS  at a small, but finite temperature,
       obtained numerically from Eq.~\ref{eq:gapequation3}.}
       \label{fig:saddle}
\end{figure}

In the weak coupling limit, the zero temperature critical field, for
$\tilde{q}=q/(2h^{1/b}) \le 1$, can be obtained from the
equation
$
          h_{c,0}(\tilde{q}) =  [2gG(\tilde{q})]^{1 \over \alpha}
$
with
\begin{equation}
          G(\tilde{q})= \int_0^\pi {d\theta \over \pi} \int_{(1-\tilde{q} \cos \theta)^b}^\infty
          d\xi  { 2 \xi^{-1+{1 \over b}} \over (\xi^{1 \over b} + \tilde{q} \cos \theta)^b +
          (\xi^{1 \over b} - \tilde{q} \cos \theta)^b}
\end{equation}
Note that now the blocking region is no longer $\xi=h$ but $\xi=h (1-\tilde{q} \cos \theta)^b$.
$G(x)$ has a maximum for $x=1$ and therefore the maximum $H_c(q)$
is reached when $\tilde{q}=q/(2h^{1/b})=1$. This field is significantly enhanced
by the VHS since it is proportional to $ V^{1 \over \alpha}$, with $\alpha=1-{1 \over b}$.
However, there is only a weak enhancement in comparison with the
$q=0$ supercooling field as observed in Fig.~\ref{fig:qual} in the case of $\alpha=1/2$.
The FF phase boundary lies clearly below the first order critical field $h_1$
and therefore, this phase will not be observed.

The same conclusion is reached if now one considers a  energy dispersion with saddle points,
$
     \xi_{\bm k}=\epsilon_{\bm k}-\epsilon_{vh}=a (\vert q_x \vert^n-\vert q_y \vert^m)
$
where
${\mathbf q}={\mathbf k}-{\mathbf k^{vh}}$, with
the values of the momentum restricted to a small
region around the saddle points ${\mathbf k^{vh}}$ and $-{\mathbf k^{vh}}$
by a cutoff. In this case, the zero gap pairing susceptibility
(the integrand of Eq.~\ref{eq:gapequation3})
has its maximum value for $q=0$ even for a simple quadratic saddle point.
In Fig.~\ref{fig:saddle}, plots of the
zero gap pairing susceptibility at zero temperature and for a fixed magnetic field are displayed
for a BCS superconductor and for an extended saddle point with $m=n=4$.
The direction of momentum ${\bm q}$ for the latter is chosen to be along the x axis or
along the diagonal of the Brillouin zone.
For the saddle point, the maximum of the pairing susceptibility is for $q=0$.
In the case of a simple quadratic saddle point, the maximum remains at $q=0$.
If one ``weakens'' the saddle point by choosing exponents smaller than two,
the maximum shifts to finite $q$,
and rapidly becomes fixed at $\tilde{q}=1$.
One could therefore conclude that the Fulde-Ferrel state is absent in systems with
energy dispersions containing saddle points close to the Fermi level.
However, one should be aware that the Fulde-Ferrel state is extremely sensitive to nesting properties
of the Fermi surface and that we have only consider the saddle point contribution to the formation of this
state and ignored the rest of the Fermi surface. In the case of homogeneous superconductivity,
such procedure is justified since the saddle point contribution clearly dominates.
In the case of Fulde-Ferrel superconductivity, if some portion of the Fermi surface
is strongly nested, it may lead to a higher critical field than that of homogeneous
superconductivity.

Concerning the experimental relevance of the results presented in this paper
to the particular case of the cuprates, an obvious statement is  that an
huge metastability region in the phase diagram is reflected by a
low temperature hysteretic behavior which can be probed, for instance, by resistive critical field measurements \cite{rcf}
or tunnelling measurements of density of states (as recently in thin Al films \cite{prl-al}).
Unfortunately, in-plane critical fields of the high-$T_c$ superconductors are presently
outside the experimental magnetic field range. One can partially
circumvent this difficulty by considering strongly overdoped or underdoped cuprates
with lower critical fields. A study of hysteresis in these materials would
provide a test of the validity of the extended Van Hove scenario for  High T$_c$ cuprate
superconductivity.

In conclusion, we have shown that a two-dimensional superconductor
with an extended saddle point in the energy dispersion pinned at
the Fermi level has at low temperature a large metastability region
in the temperature-parallel magnetic field phase diagram and that the Pauli limit, $H_p$,
for the upper critical magnetic field
is strongly enhanced in this extended Van Hove scenario.
Fulde-Ferrel superconductivity is absent from the phase diagram
unless there are Fermi surface sections (away from the saddle point region)
with very good nesting properties.




\begin{references}
\bibitem{remark} Note that the fact apparently $h_{sc,0}$ diverges as $\alpha \rightarrow 0$
just reflects the fact that in this limit the upper limit for energy in the integrals can not be extended
to infinity. The correct expression in this limit is
$h_{sc,0} = (\omega^{-\alpha} + {\alpha \over 2g})^{-1/ \alpha}$.
\end{references}
\end{document}